\documentclass[aps,twocolumn,prl,amsmath,amssymb,showpacs,floatfix,altaffilletter]{revtex4-1}
\usepackage{graphicx}
\usepackage{times}
\usepackage[varg]{txfonts}
\usepackage{bm}
\usepackage{sansmath}
\usepackage{braket}
\usepackage{xspace}
\usepackage{xr}
\usepackage{natbib} 
\usepackage{xcolor}
\usepackage[caption=false]{subfig}

\usepackage[colorlinks,bookmarks=false,citecolor=blue,linkcolor=red,urlcolor=blue]{hyperref}

\newcommand{\beq}{\begin{equation}}
\newcommand{\eeq}{\end{equation}}

\renewcommand{\vec}[1]{\boldsymbol{#1}}
\newcommand{\mat}[1]{\vec{#1}}

\begin{document}

\title{Topological Magnons in Kitaev Magnets at High Fields}

\author{P.~A.\ McClarty$^{1}$, X.-Y. Dong$^{1}$, M. Gohlke$^{1}$, J.~G.\ Rau$^{1}$, F. Pollmann$^{2}$, R. Moessner$^{1}$ and K. Penc$^{1,3}$}

\affiliation{$^{1}$ Max Planck Institute for the Physics of Complex Systems, N\"{o}thnitzer Str. 38, 01187 Dresden}
\affiliation{$^{2}$ Physics Department, Technical University Munich, James-Franck-Str. 1, 85748 Garching}
\affiliation{$^{3}$ Institute for Solid State Physics and Optics, Wigner RCP, P.O.B. 49, H-1525 Budapest, Hungary}


\begin{abstract}
We study the Kitaev-Heisenberg-$\Gamma$-$\Gamma'$ model that describes the magnetism in strong spin-orbit coupled honeycomb lattice Mott insulators. In strong $[111]$ magnetic fields that bring the system into the fully polarized paramagnetic phase, we find that the spin wave bands carry nontrivial Chern numbers over large regions of the phase diagram implying the presence of chiral magnon edge states. In contrast to other topological magnon systems, the topological nontriviality of these systems results from the presence of magnon number non-conserving terms in the Hamiltonian. Since the effects of interactions are suppressed by $J/h$, the validity of the single particle picture is tunable making paramagnetic phases particularly suitable for the exploration of this physics. Using time dependent DMRG and interacting spin wave theory, we demonstrate the presence of the chiral edge mode and its evolution with field.
\end{abstract}

\maketitle


There have been few ideas more fertile in recent condensed matter physics than the notion that band structures in solids may carry nontrivial topological indices which determine and protect certain properties of the spectrum of the solid at interfaces \cite{bernevig2013topological,hasan2010colloquium}. The core idea, formulated in the context of the integer quantum Hall effect, has led to a proliferation of novel topological states of matter including topological insulators protected by time reversal or by crystalline symmetries, as well as Weyl and Dirac semi-metals \cite{yan2017topological,armitage2017weyl} many of which have been realized in the laboratory. Analogues of this physics have been explored in photonic crystals \cite{lu2014topological}, in the mechanical properties of metamaterials \cite{mousavi2015topologically} and, even in atmospheric physics \cite{Delplace1075}. 

The concepts underlying electronic topological insulators have potentially very interesting ramifications for our understanding of magnetic materials. For example, sharp magnon bands where they exist in two dimensional ordered magnets may carry nonzero Chern number with the consequence that there are topologically protected spin waves at the edge of the system with a net chirality. A handful of models have been proposed that realize such Chern bands \cite{shindou2013topological,mook2014magnon,mook2014edge,owerre2016first,owerre2017noncollinear,romhanyi2015hall,mcclarty2017topological,nakata2017magnonic}. There is experimental evidence that such models may be realized in real materials \cite{chisnell2015topological,mcclarty2017topological}. What has been lacking on the theoretical side is a clear demonstration that the chiral edge states can be robust to the presence of interactions between magnons.

In particular, one important feature that distinguishes electronic topological insulators from their bosonic analogues is that, in the latter, interactions are more likely to play an important role possibly resulting in a breakdown of the single-particle picture. In the case of the kagome ferromagnet with Dzyaloshinskii-Moriya, it has been argued that magnon-magnon interactions broaden the bulk bands on a scale comparable to the bulk gap so that the band topology cannot be understood in terms of single magnons \cite{chernyshev2016damped}. So the question remains open whether any model can be found in which the prediction of chiral edge modes in a magnonic band structure survives in the strong coupling limit. 

In this paper, we propose a novel route to realizing topological magnon bands in systems of considerable current interest: honeycomb magnets with a significant Kitaev exchange \cite{0953-8984-29-49-493002,rau2016spin,trebst2017kitaev,hermanns2017physics,jackeli2009mott,chaloupka2010kitaev,choi2012spin,singh2010antiferromagnetic,singh2012relevance,reuther2011finite,ye2012direct,plumb2014alpha,katukuri2014kitaev,majumder2015anisotropic,chun2015direct,sears2015magnetic,PhysRevB.96.064430,banerjee2016proximate,winter2016challenges,williams2016incommensurate,wolter2017field,ponomaryov2017unconventional,winter2017breakdown,baek2017evidence,sears2017phase} some of which may be proximate to quantum spin liquid phases \cite{kitaev2006anyons,banerjee2016proximate}. The model we study has nonvanishing anomalous (number non-conserving) terms in the quadratic spin wave Hamiltonian and, in contrast to previous models of topological magnons, it is these terms that are responsible for opening up a gap in the spectrum leading to Chern bands. In addition, we present evidence that the chiral surface states that are present and topologically protected in linear spin wave theory survive the presence of magnon-magnon interactions and hence should be experimentally detectable in principle. The key to accessing this is to field-tune the system into the paramagnetic phase so that multi-magnon states are pushed to energies much higher than the single magnon states. Our time dependent density matrix renormalization group (DMRG) results provide a nonperturbative demonstration of the robustness of the chiral edge mode

\begin{figure}
\begin{center}
\includegraphics[width=\columnwidth]{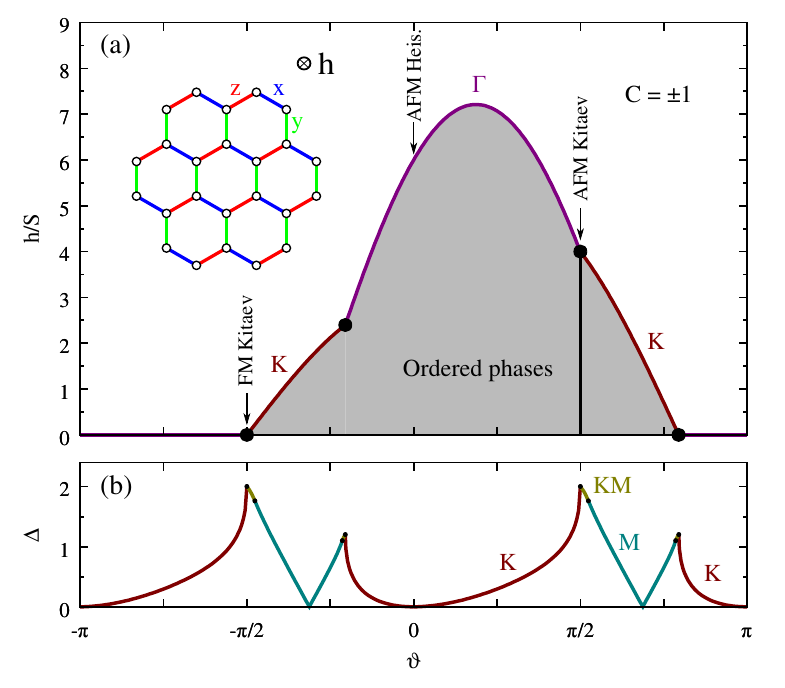}
\end{center}
\caption{
(a) The phase diagram of the Kitaev-Heisenberg model as a function of $\vartheta$ and $h/S$ as extracted from the spin wave spectrum. The region of ordered phases is determined from the gap closure in the spin wave spectrum while at the boundary of this region the corresponding ordering wavevector is indicated. The true semi-classical phase diagram has small regions, away from the Kitaev points, separated from the paramagnetic phase by first order transitions and above the threshold field shown here which are not captured using our technique \cite{PhysRevB.96.064430}. The rest of the phase diagram lies in the fully polarized phase. The entirety of the polarized paramagnetic region hosts topologically nontrivial magnons at the semiclassical level. 
Inset: honeycomb lattice cluster with $24$ sites. The different colored bonds correspond to the three types of coupling in the Kitaev model $\mathsf{S}_i^{\gamma}\mathsf{S}_j^{\gamma}$ for $\gamma=x$ (blue), $y$ (green) and $z$ (red) corresponding to the projections of the cubic axes onto the honeycomb plane. The $[111]$ field direction, indicated on the figure, is perpendicular to the plane. The exact diagonalization results presented in this paper were obtained for the Kitaev-Heisenberg model defined on this cluster. 
(b) This panel shows the minimal gap between the spin wave modes. The wavevector at which the gap is minimal is indicated by the color. 
} \label{fig:1}
\end{figure}

{\it Model} $-$ We consider the Hamiltonian \cite{rau2014generic,katukuri2014kitaev}
\begin{align}
\mathcal{H} & = J \sum_{\langle i,j \rangle } \boldsymbol{\mathsf{S}}_i\cdot \boldsymbol{\mathsf{S}}_j  \nonumber
\\ & + \sum_{\langle i,j \rangle_{\gamma}} \left\{ 2K \mathsf{S}_{i}^\gamma \mathsf{S}_{j}^\gamma + \Gamma \left(\mathsf{S}_{i}^{\alpha}\mathsf{S}_{j}^{\beta} + \mathsf{S}_{i}^{\beta}\mathsf{S}_{j}^{\alpha} \right) \right\} - \boldsymbol{h} \cdot \sum_{i} \boldsymbol{\mathsf{S}}_i
\label{eq:HJKG}
\end{align}
where the indices $\{\alpha,\beta,\gamma\}$ run over components $\{x,y,z\}$ with the $\gamma$ component corresponding to one of the three types of bond as indicated in Figure~\ref{fig:1}(a). Models with significant $K/J$ has been proposed to underlie the correlated magnetism observed in the effective spin one-half systems A$_2$IrO$_3$ (A=Na,Li) \cite{singh2010antiferromagnetic,singh2012relevance,choi2012spin,rau2014generic,katukuri2014kitaev,chun2015direct,williams2016incommensurate} and $\alpha$-RuCl$_3$ \cite{plumb2014alpha,banerjee2016proximate,sears2015magnetic,winter2017breakdown,little2017antiferromagnetic,wang2017magnetic,ran2017spin} following theoretical work that laid the basis for the possibility of large compass-type interactions in such honeycomb magnets \cite{jackeli2009mott,chaloupka2010kitaev}. A fourth exchange coupling $\Gamma'$ is allowed by symmetry \cite{SM,rau2014generic,katukuri2014kitaev}. We postpone discussion of the effects of the $\Gamma$ and $\Gamma'$ terms until later in the paper and focus, for now, on the remaining Hamiltonian. We parameterize this Kitaev-Heisenberg model using angle $\vartheta$ so that $J=\cos\vartheta$ and $K=\sin\vartheta$. 

From now on, we consider the case where the magnetic field, of magnitude $h$, is applied parallel to $[111]$ (Fig.~\ref{fig:1}). For  $h$ greater than some threshold, the moments are fully polarized in the field direction and we expand the moments in small fluctuations about this collinear state in Holstein-Primakoff bosons \cite{holstein1940field}. The quadratic Hamiltonian that results at order $S$ is of the form  $\mathcal{H}_{\rm KH-LSW}  = \sum_{\boldsymbol{k}} \boldsymbol{\Upsilon}_{\boldsymbol{k}}^\dagger \boldsymbol{\mathsf{M}}(\boldsymbol{k})  \boldsymbol{\Upsilon}_{\boldsymbol{k}}$
where $\boldsymbol{\Upsilon}_{\boldsymbol{k}}  = \left( a_{\boldsymbol{k}}^{} \ b_{\boldsymbol{k}}^{} \ a^\dagger_{-\boldsymbol{k}} \ b^\dagger_{-\boldsymbol{k}}  \right)^T$ with $a$ and $b$ bosons living on the two different honeycomb sublattices. The $4\times 4$ Hamiltonian $\boldsymbol{\mathsf{M}}(\boldsymbol{k})$, which is given explicitly in the Supplementary Material, takes the form
\begin{align}
 \boldsymbol{\mathsf{M}}(\boldsymbol{k})  = \left( \begin{array}{cc}   
\mathsf{A}(\boldsymbol{k}) & \mathsf{B}(\boldsymbol{k}) \\ \mathsf{B}^{\dagger}(\boldsymbol{k}) & \mathsf{A}^{T}(-\boldsymbol{k})
\end{array} \right),
\end{align}
where the $\mathsf{A}$ block contains the number conserving terms $a^\dagger a$ and $\mathsf{B}$ contains the number non-conserving terms $a^\dagger a^\dagger$. The eigenproblem for this Hamiltonian, leading to two spin wave branches $\omega^{\alpha}_{\boldsymbol{k}}$ for $\alpha=1,2$, may be solved by performing a bosonic Bogoliubov transformation.

The phase diagram of the $J-K$ model is shown in Fig.~\ref{fig:1}(a) indicating the fully polarized phase and regions of spontaneous magnetic order obtained by finding the couplings at which magnons condense - the translational symmetry of these phases is then determined by the condensation wavevector. The precise nature of the ordered states can be found in Ref.~\onlinecite{PhysRevB.96.064430}. 


\begin{figure*}
\begin{center}
\includegraphics[width=\textwidth]{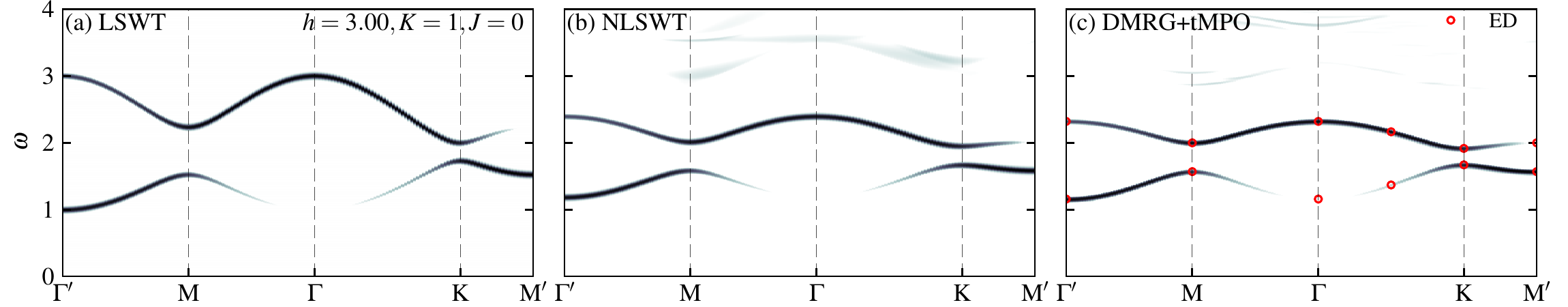}
\caption{Dynamical correlation function $S(\boldsymbol{k},\omega)$ computed along high symmetry lines for $h=3$ ($S=1/2$) at the AFM Kitaev point $\vartheta=\pi/2$. The intensity scale is logarithmic from $5\times 10^-3$ to $1$. (a) Linear spin wave theory, (b) nonlinear spin wave theory and (c) time dependent DMRG all with gaussian broadening of the lines for purposes of presentation $\sigma_{\omega}=0.01$. The overlaid points in (c) show single magnon states obtained from $24$ site ED with periodic boundary conditions.} \label{fig:2}
\end{center}
\end{figure*}

{\it Non-interacting Magnons.} $-$ We first focus on the antiferromagnetic Kitaev point $\vartheta=\pi/2$ ($J=0$). The linear spin wave dispersions along high symmetry lines are shown in Fig.~\ref{fig:2}(a) for $h/S=6$. For $h/S>4$, the spectrum exhibits both a nonzero gap to the lowest mode and a gap between the modes. As $h/S\rightarrow 4$ the lowest mode falls to zero across the entire zone, corresponding to the onset of a classical spin liquid regime, while the highest mode is completely gapped and dispersive. 

Since the two bands do not touch at $\vartheta=\pi/2$, the Berry curvature is everywhere well defined (see Supp. Mat. \cite{SM}). The Chern numbers of the two bands at this coupling are $\pm 1$ for all $h/S\geq 4$ implying the existence of chiral magnon edge modes. Since the bulk topology is not altered by modifications to the ground state in the vicinity of the boundary, we may illustrate the phenomenon with a linear spin wave calculation of the spectrum above the collinear spin state on a slab geometry with zigzag boundaries parametrized by the momentum along the translation invariant direction. Such a calculation \cite{SM} reveals a pair of modes, each with a well-defined chirality, running between the bulk bands and with the weight of the wavefunction of these interband modes concentrated at opposite edges. 

We may gain some insight into the mechanism that leads to the topological magnon bands. The $\vartheta=0$ Hamiltonian ($K=0$) has only number conserving terms and the two magnon bands meet at the $K$ point. Inspection of the Hamiltonian $\boldsymbol{\mathsf{M}}(\boldsymbol{k})$ shows that the $\mathsf{A}$ block contains the two couplings only in the combination $J + \frac{2K}{3}$ so the number non-conserving terms of the $\mathsf{B}$ block must be responsible for the gap opening between the bands in the magnon spectrum. These terms also break an effective time reversal symmetry leading to the identification of the magnon bands with the topological insulator class D \footnote{Physical time reversal is broken both by spontaneous magnetic order and by the external magnetic field.  For our purposes one must ask whether there is some antiunitary operator $\mathcal{T}\equiv U_{T}K$ that acts on the Hamiltonian such that $U_T^\dagger \boldsymbol{\mathsf{M}}^{\star}(\boldsymbol{k})U_T=\boldsymbol{\mathsf{M}}(-\boldsymbol{k})$. For the $\vartheta=0$ model the unitary operator $U_T$ acts trivially and the tight-binding model has an effective time reversal symmetry. This is broken by the anomalous terms when the Kitaev term is switched on.}. 

The observation that the gap closes as $1/h$ at high fields suggests that further insight may be gained by carrying out a Schrieffer-Wolff transformation perturbatively in the anomalous terms to obtain an effective Hamiltonian in the number-conserving sector. One finds to second order that the nearest neighbor coupling is renormalized and effective second neighbor hopping terms are generated that are of the same form as those arising from a bare second-neighbor Dzyaloshinskii-Moriya exchange coupling. In short, at very high fields, the spin wave spectrum of the Kitaev model reduces to that of the honeycomb ferromagnet with second neighbor DM exchange that is known from earlier work to exhibit Chern bands \cite{owerre2016first}.

We now consider the entire $J-K$ semiclassical paramagnetic regime. The lower panel of Fig.~\ref{fig:1}(b) shows that the two magnon bands touch at four distinct $\vartheta$ including $\vartheta=0,\pi$ and are otherwise gapped. Away from these lines in $\vartheta - h$, the magnon bands are topologically nontrivial. We further note \cite{SM} that the spin wave spectrum at some $\vartheta$ and field $(h-h_{\rm th}(\vartheta))/S$ is identical to the spectrum at $\vartheta+\pi$ and $(h-h_{\rm th}(\vartheta+\pi))/S$ where $h_{\rm th}(\vartheta)$ is the threshold field. The band topology is preserved by the mapping so, for example, the ferromagnetic Kitaev point with zero semiclassical threshold field has Chern magnon bands following from results at the $\vartheta=\pi/2$ point.

Finally, to make contact with materials, we observe that in the full $J-K-\Gamma-\Gamma'$ nearest neighbor model of Eq.~\ref{eq:HJKG}  and Ref.~\onlinecite{SM} in the fully polarized phase, the linear spin wave Hamiltonian is related to the Kitaev-Heisenberg model through a mapping of the parameters $J\rightarrow J-\Gamma$, $K\rightarrow K+\Gamma$ and $h\rightarrow h-3\Gamma S$ so topological magnon bands are expected to be present in Kitaev magnets in the paramagnetic regime at least where spin wave interactions may be neglected.  


{\it Beyond Linear Spin Wave Theory.} $-$ By expanding in Holstein-Primakoff bosons to order $O(1/S^2)$, one finds three-boson and four boson terms in the Hamiltonian. The former arise in an expansion around the collinear ground state owing to the anisotropic nature of the exchange. Both the three-body and a set of four-body couplings violate particle number conservation and provide a mechanism for the magnons to acquire a finite lifetime. Upon lowering the field the two magnon states eventually overlap with the single magnon states so that one to two magnon decay is kinematically allowed leading to broadening of the single magnon modes. This process may also lead to the destruction of the chiral edge mode if the widths of the bulk bands or that of the edge mode become comparable to the gap between the magnon bands. 

To address the effect of interactions on the bulk spectrum and chiral edge mode we extend the analysis of the previous section in three ways (i) perturbatively in the magnon-magnon interactions to $O(1/S^2)$ in spin wave theory (NLSWT) \cite{blaizot1986quantum,chubukov1994large,chernyshev2009triangular,mourigal2013dynamical,chernyshev2013colloquium} and (ii) nonperturbatively using DMRG with a matrix product operator based time evolution (DMRG + tMPO) \cite{PhysRevLett.119.157203,PhysRevB.91.165112} and (iii) with exact diagonalization (ED) of the Hamiltonian on a $24$ site cluster that preserves the lattice symmetries. 

First we examine the dynamical correlation function 
\begin{equation} 
S(\boldsymbol{k},\omega)\equiv \sum_{\alpha} S^{\alpha\alpha}(\boldsymbol{k},\omega) =  \sum_{\alpha}\sum_{a,b}  \langle  \mathsf{S}_a^{\alpha}(-\boldsymbol{k},-\omega) \mathsf{S}_b^{\alpha}(\boldsymbol{k},\omega) \rangle,
\end{equation}
at the $\vartheta=\pi/2$ point for various fields using LSWT, NLSWT and DMRG+tMPO. For the latter, the calculations were performed on infinite cylinders with a circumference of $8$ sites ($L_x =4$) by a MPO based time evolution of the wavefunction after a single spin flip is performed on the ground state wavefunction.  Results at $h=3$ are shown in Fig.~\ref{fig:1}(a)-(c). The                                                                                             apparently well-defined. The upper mode has only a small dispersion at this field and the continuum has a low intensity. The gap closes only at $h \approx 1.25$ ($S=1/2$). The supplementary section shows corresponding plots for the ferromagnetic Kitaev point, $\vartheta=3\pi/2$ \cite{SM}.

To address the fate of the chiral edge modes that are topologically protected within LSWT, we show DMRG+tMPO and NLSWT results for a slab geometry with one periodic direction and two open boundaries. Since the introduction of a boundary destabilizes the fully polarized spin configuration in the vicinity of the edge, LSWT and NLSWT results were obtained by first solving for the non-collinear classical ground state on the slab and perturbing about this solution. All results were obtained for a slab periodic in $y$ with dimensions $L_{\rm x}=5$ unit cells and length $L_{\rm y}=71$ for the DMRG chosen to ensure that long enough times could be reached for the requisite energy resolution without entanglement spreading to the $y$ boundaries of the slab. 

\begin{figure}
\begin{center}
\includegraphics[width=\columnwidth]{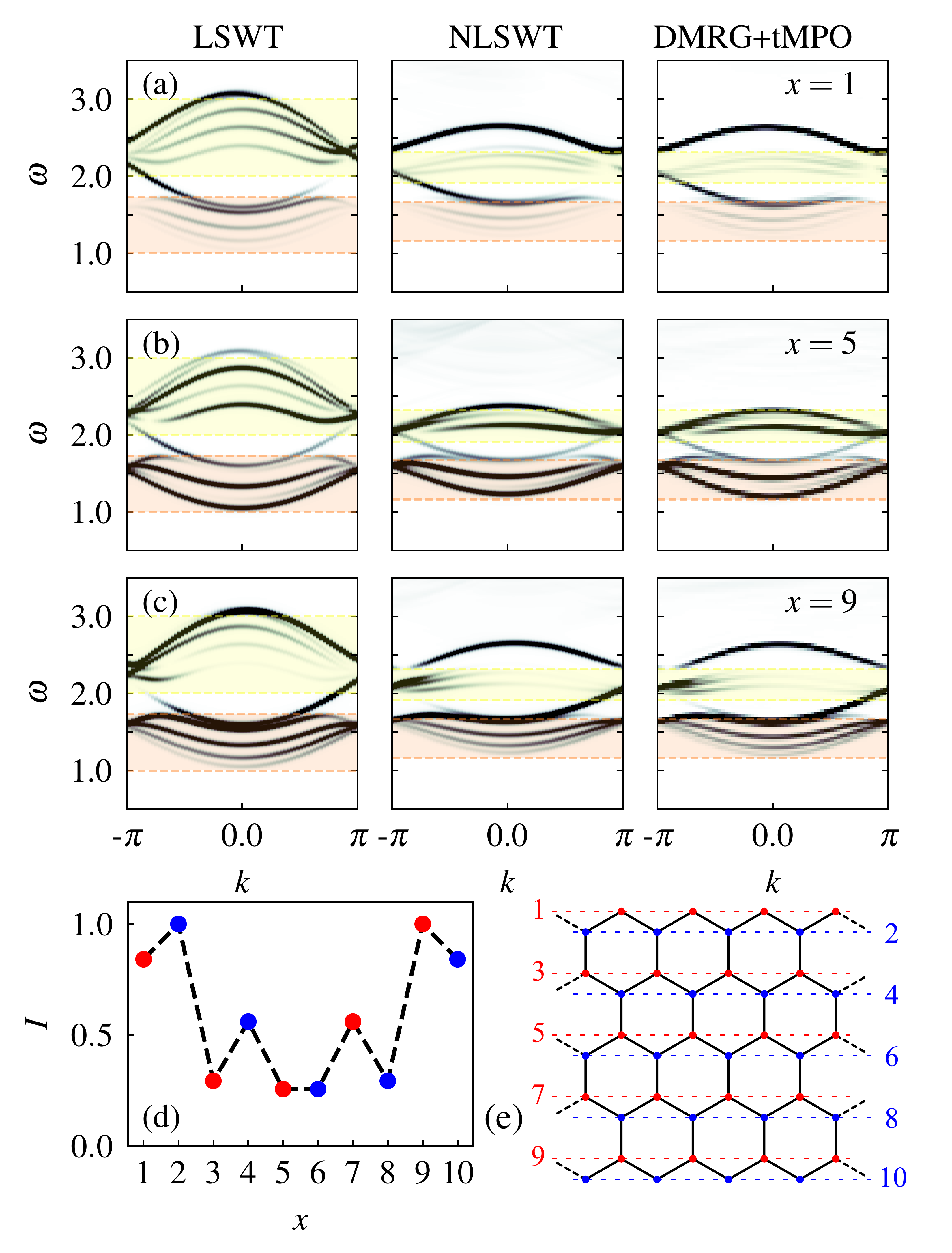}
\caption{Dynamical correlation functions $ \langle  \mathsf{S}^{x}(-k,x,-\omega) \mathsf{S}^{x}(k,x,\omega)\rangle$ computed using DMRG+tMPO (right column) on a slab [illustrated in panel (e)] of width $L_{\rm x}=5$ unit cells and length $L_{\rm y}=71$ and with periodic boundary conditions imposed along $y$. The crystal momentum along the translationally invariant direction is denoted by $k$ and $x$ is the line number [indicated in (e)]. The DMRG+tMPO energy resolution is $\Delta\omega\approx 0.03$. For comparison, corresponding results are shown for linear spin wave theory (left column) and $O(S^0)$ nonlinear spin wave theory (middle column). All calculations were performed for $\vartheta=\pi/2$ and $h=3$. The plots in each row correspond to different line numbers (from top to bottom $x=1, 5$ and $9$). The shading in each figure indicates the bulk single magnon band widths with the bulk band gap in between the shaded bands. The chiral modes at the two boundaries $x=1$ (top) and $x=9$ (bottom) and their edge character can been seen through the reduction in the band gap intensity in the middle of the slab $x=5$ (middle) and in panel (d) which shows the integrated intensity within the bulk band gap in different layers on the slab. The velocities of the edge modes are opposite at the two boundaries.} \label{fig:3}
\end{center}
\end{figure}

Fig.~\ref{fig:3} illustrates dynamical correlations on the slab for the different methods introduced above. The slab geometry is shown in panel (e) and the different rows $(a)$ to $(c)$ show the $k$ dependent correlations on different lines through the slab including the two edges. LSWT for this geometry [Fig.~\ref{fig:3}(left column)] shows that the slab is thick enough for the chiral edge modes (which have opposite directions for the two edges) to be well resolved. Fig.~\ref{fig:3} further shows that the chiral edge modes survive in the full nonperturbative interacting spin model (right) albeit with significant renormalization of the bulk modes which is almost entirely captured by the interacting spin wave calculation (middle). Panel (d) shows quantitatively that the intensity between the bulk bands is concentrated at the edges.

\begin{figure}[tp]
\begin{centering}
\vspace{20pt}
\includegraphics[width=0.9\columnwidth]{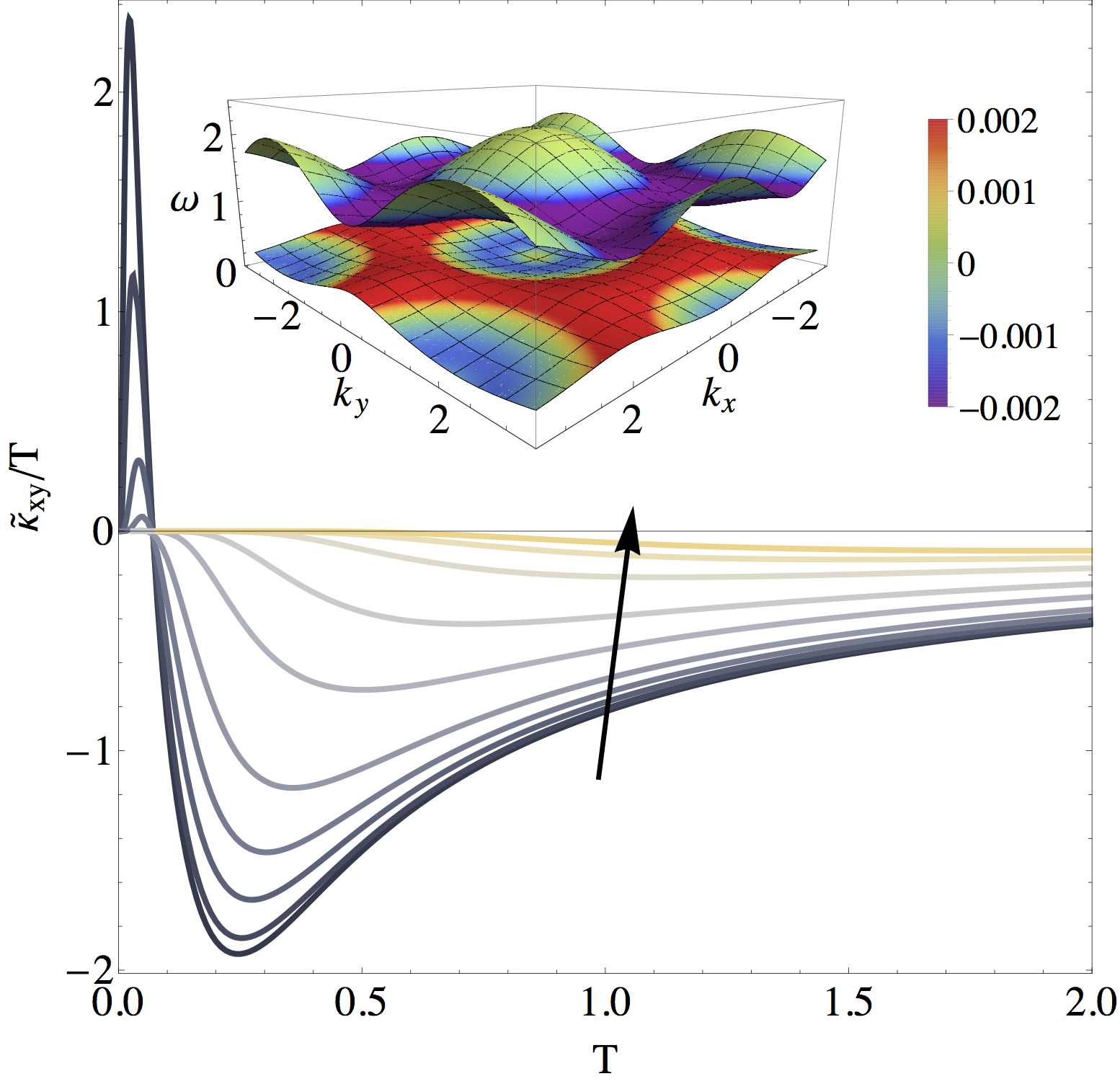}
\caption{Dimensionless thermal Hall conductivity $\tilde{\kappa}_{xy}/T$ as a function of temperature at the ferromagnetic Kitaev point ($S=1/2$), $K=-1$, at fields $h=0.01,0.02,0.05,0.1,0.2,0.5,1,2,3,4$ to be read in the arrow direction. The inset shows the dispersions of the bands at $h=0.1$. The color of the bands indicates the Berry curvature with the scale corresponding to the color bar.} \label{fig:4}
\end{centering}
\end{figure}

{\it Thermal Hall Effect.} $-$ The presence of a nontrivial Berry curvature in the magnon bands implies the existence of a thermal Hall signature provided that the Berry curvature is not odd in momentum. The magnon thermal Hall effect has been investigated both theoretically and experimentally in a number of magnets \cite{katsura2010theory,matsumoto2011rotational,matsumoto2014thermal,mook2014magnon,romhanyi2015hall,onose2010observation,ideue2012effect,murakami2016thermal,Watanabe8653}. Earlier theoretical work has explored the thermal Hall response at low fields \cite{nasu2017thermal} in the Kitaev honeycomb model. Here we extend the analysis to Kitaev systems in the high field regime. Fig.~\ref{fig:4}(b) shows the dimensionless thermal Hall conductivity $\tilde{\kappa}_{xy}/T$ \cite{SM} as a function of temperature and for various magnetic fields at $\vartheta=3\pi/2$. The shape of the function $\tilde{\kappa}_{xy}/T$ can be understood as follows. As a function of field, the magnon bands are gapped out resulting in an exponential decrease in the thermal Hall signature in $h$. The sign change in $\tilde{\kappa}_{xy}$ at low temperatures and fields reflects the variation in the Berry curvature in momentum space - the Berry curvature is positive in the vicinity of $\Gamma$ in the lowest band, this being is the maximally thermally occupied state at very low temperatures, while it changes sign for larger momenta. As $T \rightarrow \infty$, $\tilde{\kappa}_{xy}$ saturates to a constant value. It will be interesting to examine recent thermal Hall signatures in $\alpha$-RuCl$_3$ in the light of these results \cite{kasahara2017thermal}. 
 

{\it Discussion.} $-$ The Hilbert space of a bosonic system encompasses an infinite tower of multiparticle excitations. In insulating magnets, the single magnon sector is meaningful only when the number non-conserving terms in the magnon Hamiltonian are suppressed, for example, by symmetry, in powers of $1/S$, or when there is a separation of energy scales between the many-magnon states. It is the latter case that we have advocated for in this paper as a means of exploring topologically protected magnon edge states. By field-tuning the magnet into a paramagnetic phase, the single magnon branches are gapped out and linewidths are suppressed. We have studied the concrete case of spin-orbit coupled honeycomb magnets with significant $K$, $J$, $\Gamma$ and $\Gamma'$ terms finding that the paramagnetic regime generically has Chern magnon bands with chiral edge states. The topological nontriviality is enforced by the anomalous terms in the quadratic Hamiltonian. We anticipate that future experimental developments will facilitate direct measurements of the edge states in such systems.

\begin{acknowledgments}
K.P. acknowledges support from MPI PKS for an extended stay at the institute during which part of this work was completed. This
work was in part supported by the Deutsche Forschungsgemeinschaft under grant SFB 1143. 
\end{acknowledgments}

\bibliography{references}

\onecolumngrid

\section{Supplementary Material for `Topological Magnons in the Kitaev Honeycomb Model at High Fields'}

\section{Symmetry-allowed Exchange Hamiltonian}

Here we will derive the symmetry allowed terms for the edge-shared octahedral compounds (see also Ref.~\onlinecite{katukuri2014kitaev,rau2014generic}). The Ir$^{4+}$ and Ru$^{3+}$ in the center of oxygen octahedra forms a honeycomb lattice. Taking into account the octahedral environment, the symmetry group of a single bond 
consists of the inversion, the two-fold rotation around the bond, and their composition, a mirror plane which is perpendicular to the bond. The point group at the center of the hexagon is $D_{3d}$, generated by the $S_6$ rotoreflection (we note that $S_6^3$ is the inversion) and the mirror plane.  The point group at the Ir site is $D_3$, with a $C_3$ threefold rotation and a $C'_2$ rotation as generators.

\begin{figure}[htbp]
\begin{center}
\includegraphics[width=0.55\columnwidth]{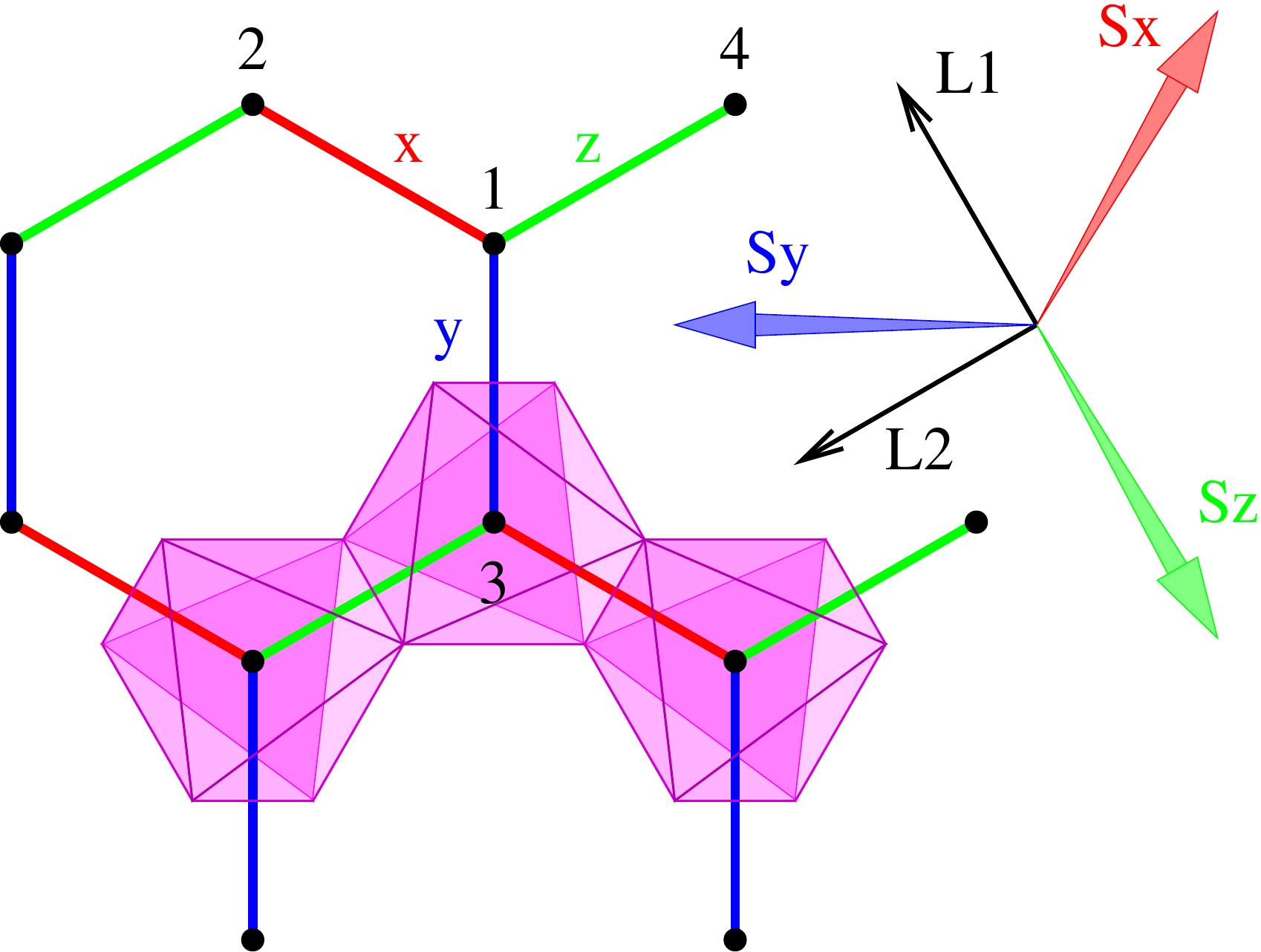}
\caption{Schematic of the lattice structure: the magnetic sites (black circles) in octahedral environment (magenta) form a honeycomb lattice. The $x$ (red bonds), $y$ (blue), and $z$ (green) Kitaev interactions are indicated. The spin components $\mathsf{S}^x$, $\mathsf{S}^y$, and $\mathsf{S}^z$ form an orthogonal bases, shown on the right. $L^{(1)}$ and $L^{(2)}$ are the two components of rotated spins used to derive the spin wave Hamiltonian, when the $L^{(3)}$ is perpendicular to the honeycomb plane and parallel to the external field direction $[111]$. The numbers 1,2,3, and 4 denote lattice sites used in the text.
}
\label{fig:axes}
\label{axes} 
\end{center}
\end{figure}

First, let us consider the $x$ bond in the geometry presented in Fig.~\ref{axes}. The inversion at the center of the bond exchanges the two sites without affecting the spin components, $\mathsf{S}^\alpha_1 \leftrightarrow \mathsf{S}^\alpha_2$. The $C_2$ rotation does not exchange the sites; it acts only on the spin components as $\mathsf{S}^x \to -\mathsf{S}^x$ [the $\mathsf{S}^\alpha$ component of the spin is perpendicular to the $\alpha=x,y,z$ bond and, with the constraint that the components form an orthogonal basis, there are eight possible choices of axis convention out of which we have chosen one (Fig.~\ref{fig:axes})],  $\mathsf{S}^y \to -\mathsf{S}^z$, and $\mathsf{S}^z \to -\mathsf{S}^y$. 
We can construct the following 4 invariants:
\begin{align}
& \mathsf{S}^x_1 \mathsf{S}^x_2,\\
& \mathsf{S}^y_1 \mathsf{S}^y_2 + \mathsf{S}^z_1 \mathsf{S}^z_2 \\
& \mathsf{S}^z_1 \mathsf{S}^y_2 + \mathsf{S}^y_1 \mathsf{S}^z_2, \\
& \mathsf{S}^x_1 \mathsf{S}^y_2 + \mathsf{S}^x_1 \mathsf{S}^z_2 + \mathsf{S}^y_1 \mathsf{S}^x_2 + \mathsf{S}^z_1 \mathsf{S}^x_2 .
\end{align}
Adding them up with a suitable coefficients we arrive to 
\begin{align}
  \mathcal{H}_x &= 2 K \mathsf{S}^x_1 \mathsf{S}^x_2 + J \mathbf{\mathsf{S}}_1 \cdot \mathbf{\mathsf{S}}_2 + \Gamma \left(\mathsf{S}^z_1 \mathsf{S}^y_2 + \mathsf{S}^y_1 \mathsf{S}^z_2\right) \nonumber\\ 
   &\phantom{=} +  \Gamma' \left( \mathsf{S}^x_1 \mathsf{S}^y_2 + \mathsf{S}^x_1 \mathsf{S}^z_2 + \mathsf{S}^y_1 \mathsf{S}^x_2 + \mathsf{S}^z_1 \mathsf{S}^x_2\right)
\end{align}
The inversion symmetry about the bond centers prevents DM interactions.

Next, we use the $C_3$ rotation about site 1 to get the Hamiltonian for the $y$ bond, as it cyclically exchanges the three spin components, $\mathsf{S}^x \to \mathsf{S}^y \to \mathsf{S}^z \to \mathsf{S}^x$, furthermore  site 1 remains, and site 2 becomes site 3:
\begin{align}
  \mathcal{H}_y &= 2 K \mathsf{S}^y_1 \mathsf{S}^y_3 + J \mathbf{\mathsf{S}}_1 \cdot \mathbf{\mathsf{S}}_3 + \Gamma \left(\mathsf{S}^x_1 \mathsf{S}^z_3 + \mathsf{S}^z_1 \mathsf{S}^x_3\right) \nonumber\\ 
   &\phantom{=} +  \Gamma' \left( \mathsf{S}^y_1 \mathsf{S}^z_3 + \mathsf{S}^y_1 \mathsf{S}^x_3 + \mathsf{S}^z_1 \mathsf{S}^y_3 + \mathsf{S}^x_1 \mathsf{S}^y_3 \right),
\end{align}
and a similar form for the $\mathcal{H}_z$. This is the Hamiltonian first established in Refs.~\onlinecite{jackeli2009mott,chaloupka2010kitaev,rau2014generic,katukuri2014kitaev}.

\section{Linear Spin Wave Theory}

The honeycomb lattice is triangular $\boldsymbol{R}_{mn}=\left( m-(n/2), \sqrt{3}n/2 \right)$ with a two site basis. We take the bonds to be
\begin{align*}
\boldsymbol{\delta}_x & = \left( 0, 1 \right), \\
\boldsymbol{\delta}_y & = \left( -\frac{\sqrt{3}}{2}, -\frac{1}{2} \right), \\
\boldsymbol{\delta}_z & = \left(  \frac{\sqrt{3}}{2}, -\frac{1}{2}  \right), \\
\end{align*}

As outlined in the main text, we can define the following vector of the bosonic operators
\begin{align} 
 \mathbf{\Upsilon}_{\boldsymbol{k}} &= ( a^{\phantom{\dagger}}_{\boldsymbol{k}} , b^{\phantom{\dagger}}_{\boldsymbol{k}} , a^\dagger_{-\boldsymbol{k}} , b^\dagger_{-\boldsymbol{k}} ) \;.
 \end{align}
Then the linear spin wave Hamiltonian can also be cast into the form
\begin{equation}
\mathcal{H}_\text{LSW} = 
\frac{1}{2} \sum_{\boldsymbol{k}\in \text{BZ}} 
\mathbf{\Upsilon}^{\dagger}_{\boldsymbol{k}} 
\cdot \mathsf{H}_\text{LSW}(\boldsymbol{k}) \cdot 
\mathbf{\Upsilon}^{\phantom{\dagger}}_{\boldsymbol{k}}
\label{eq:LSW}
\end{equation}
For the Kitaev-Heisenberg model we find \cite{PhysRevB.96.064430}
\begin{align}
 \mathsf{H}_{\rm KH-LSW}(\boldsymbol{k}) = \left( \begin{array}{cc}   
\mathsf{A}(\boldsymbol{k}) & \mathsf{B}(\boldsymbol{k}) \\ \mathsf{B}^{\dagger}(\boldsymbol{k}) & \mathsf{A}^{T}(-\boldsymbol{k})
\end{array} \right)
\end{align}
where
\begin{align}
\mathsf{A}(\boldsymbol{k}) & = \left( \begin{array}{cc}   
-3JS - 2KS + h &  \left( J + \frac{2K}{3} \right)S \left( e^{i\boldsymbol{k}\cdot\boldsymbol{\delta}_x} + e^{i\boldsymbol{k}\cdot\boldsymbol{\delta}_y} + e^{i\boldsymbol{k}\cdot\boldsymbol{\delta}_z}  \right)  \\ 
 \left( J + \frac{2K}{3} \right)S \left( e^{-i\boldsymbol{k}\cdot\boldsymbol{\delta}_x} + e^{-i\boldsymbol{k}\cdot\boldsymbol{\delta}_y} + e^{-i\boldsymbol{k}\cdot\boldsymbol{\delta}_z}  \right)& -3JS - 2KS + h
\end{array} \right) \\
\mathsf{B}(\boldsymbol{k}) & = \left( \begin{array}{cc}   
0 & \frac{2KS}{3} \left( e^{i\boldsymbol{k}\cdot\boldsymbol{\delta}_x  + \frac{2\pi i}{3}} + e^{i\boldsymbol{k}\cdot\boldsymbol{\delta}_y - \frac{2\pi i}{3}} + e^{i\boldsymbol{k}\cdot\boldsymbol{\delta}_z}  \right) \\
\frac{2KS}{3}  \left( e^{-i\boldsymbol{k}\cdot\boldsymbol{\delta}_x  + \frac{2\pi i}{3}} + e^{-i\boldsymbol{k}\cdot\boldsymbol{\delta}_y - \frac{2\pi i}{3}} + e^{-i\boldsymbol{k}\cdot\boldsymbol{\delta}_z}  \right) & 0
\end{array} \right) 
\end{align}

In the main text, we parametrized the couplings using $\vartheta$ such that $J=\cos\vartheta$ and $K=\sin\vartheta$.

It is convenient to introduce
\begin{align}
\gamma^{\phantom{*}}_{0,\boldsymbol{k}} &= \frac{1}{3} (e^{-i \boldsymbol{k}\cdot \bm{\delta}_{x}}+e^{-i \boldsymbol{k}\cdot\bm{\delta}_{y}}+e^{-i \boldsymbol{k}\cdot\bm{\delta}_{z}}),
\\
\gamma^{\phantom{*}}_{1,\boldsymbol{k}} &= \frac{1}{3} (e^{-i \boldsymbol{k}\cdot\bm{\delta}_{x}-(2 \pi i/3)}+e^{-i \boldsymbol{k}\cdot\bm{\delta}_{y}+(2 \pi i/3)}+e^{-i \boldsymbol{k}\cdot\bm{\delta}_{z}}),
\\
\gamma^{\phantom{*}}_{2,\boldsymbol{k}} &= \frac{1}{3} (e^{-i \boldsymbol{k}\cdot\bm{\delta}_{x}+(2 \pi i/3)}+e^{-i \boldsymbol{k}\cdot\bm{\delta}_{y}-(2 \pi i/3)}+e^{-i \boldsymbol{k}\cdot\bm{\delta}_{z}}),
\end{align}
satisfying the relations
$\gamma^*_{0,\boldsymbol{k}}=\gamma^{\phantom{*}}_{0,-\boldsymbol{k}}$, $\gamma^*_{1,\boldsymbol{k}}=\gamma^{\phantom{*}}_{2,-\boldsymbol{k}}$, and
$\gamma^*_{2,\boldsymbol{k}}=\gamma^{\phantom{*}}_{1,-\boldsymbol{k}}$ so that
\begin{align} 
\mathsf{A}(\boldsymbol{k}) &= h \left(
\begin{array}{cc}
 1 & 0 \\
 0 & 1 \\
\end{array}
\right)
+(3 J + 2 K) S \left(
\begin{array}{cc}
 -1 &  \gamma^*_{0,\boldsymbol{k}} \\
  \gamma^{\phantom{*}}_{0,\boldsymbol{k}} & -1 \\
\end{array}
\right) ,
\\
\mathsf{B}(\boldsymbol{k}) &= 2 K S\left(
\begin{array}{cc}
 0 & \gamma^*_{1,\boldsymbol{k}} \\
 \gamma^{\phantom{*}}_{2,\boldsymbol{k}} & 0 \\
\end{array}
\right) .
\end{align}

\subsection{Mapping $\vartheta\rightarrow\vartheta+\pi$}
\label{sec:map}

Under the mapping $\vartheta\rightarrow\vartheta+\pi$, the couplings flip sign. There is a simple relationship between the spectra of the linear spin wave Hamiltonian under this mapping when combined with a field redefinition $h\rightarrow h-6JS-4KS$, that preserves the diagonal matrix elements, and $\boldsymbol{k}\rightarrow -\boldsymbol{k}$. This whole transformation can be undone by a unitary transformation of the form
\begin{equation*}
\left( \begin{array}{cccc} 0 & 1 & 0 & 0 \\  -1 & 0 & 0 & 0 \\  0 & 0 & 0 & 1 \\  0 & 0 & -1 & 0   \end{array}  \right).
\end{equation*}
It follows that, under the mapping, as measured from the threshold field the spectrum is left unchanged.

\begin{figure}
\begin{center}
\includegraphics[width=0.65\columnwidth]{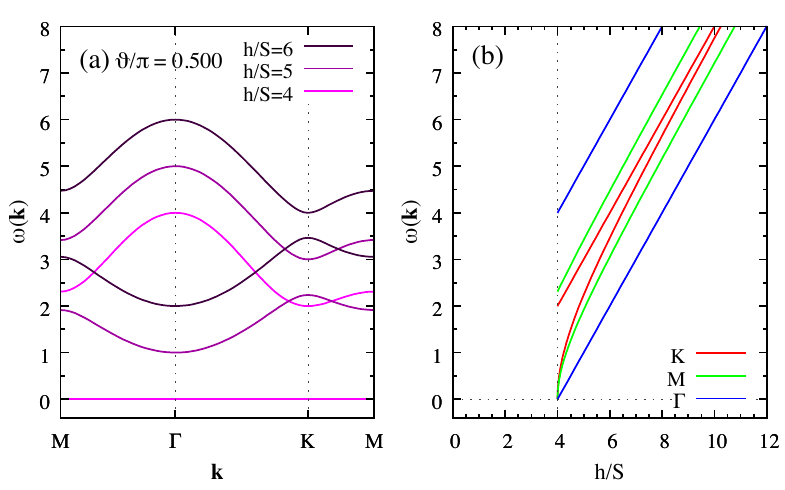}
\caption{(a) Magnon dispersion relations along high symmetry lines, as marked in the upper panel, for three different fields $h/S=4,5,6$ at the isotropic AFM point $\vartheta=0$.  Panel (b) shows the spin wave energies as a function of field at three high symmetry points, illustrating the linear dependence of the lowermost mode at the $\Gamma$, $K$ and $M$ points.} \label{fig:2}
\end{center}
\end{figure}

\subsection{The Role of the $\Gamma$ and $\Gamma'$ Terms}

The $\Gamma$ and $\Gamma'$ terms are symmetric exchange couplings that are allowed by the symmetries of the honeycomb lattice as discussed above. 

For fully polarized moments in the $[111]$ direction, the $\Gamma$ and $\Gamma'$ couplings merely effect the following mapping on the $J$, $K$, $h$ model:
\begin{subequations}
\label{equations}
 \begin{align}
 K &\to K + \Gamma - \Gamma' \;, \\
 J &\to J - \Gamma \;,  \\
 h &\to h - 3 \Gamma S - 6 \Gamma' S  \;.
\end{align} 
\end{subequations}
In other words, all the results we obtained for the Kitaev-Heisenberg model can be extended to the $JK\Gamma\Gamma'$ model using the replacement rules above.

\subsection{The canonical transformation}

In this section, we examine the linear spin wave theory at high fields and systematically integrate out the number non-conserving terms in powers of $1/h$ to obtain an effective hopping Hamiltonian for the magnons.

To proceed, we look for a canonical transformation
 \begin{align} 
\mathcal{H}_{\text{eff}} &= e^\mathcal{W} \mathcal{H} e^{-\mathcal{W}} \nonumber\\
   &= \mathcal{H} + [\mathcal{W}, \mathcal{H}] + \frac{1}{2} [\mathcal{W}, [\mathcal{W}, \mathcal{H}]] + \cdots
 \end{align}
where the operator $\mathcal{W}$ is chosen such that the transformation eliminates the magnon number non-conserving anomalous terms to $O(1/h)$ instead capturing their effect to this order in terms of a pure hopping Hamiltonian. This is achieved by choosing 
\begin{align} 
\mathcal{W} = 
\frac{K S}{h} \sum_{\boldsymbol{k}\in \text{BZ}} 
\left( 
 \gamma^*_{1,\boldsymbol{k}} a^{\dagger}_{\boldsymbol{k}} b^\dagger_{-\boldsymbol{k}} 
- \gamma^{\phantom{*}}_{1,\boldsymbol{k}} a^{\phantom{\dagger}}_{\boldsymbol{k}} b^{\phantom{\dagger}}_{-\boldsymbol{k}} 
\right),
\end{align}
or, in the matrix notation we used above (Eq.~\ref{eq:LSW}) for the linear spin wave Hamiltonian,
\begin{equation} 
\mathsf{W}(\boldsymbol{k}) 
= \frac{1}{2h} \left(
\begin{array}{cc}
 0 & \mathsf{B}(\boldsymbol{k}) \\
 -\mathsf{B}^\dagger(\boldsymbol{k}) & 0 \\
 \end{array}
\right) .
\end{equation}

The effective Hamiltonian is then characterized by the
 \begin{align} 
\mathsf{A}_{\text{eff}}(\boldsymbol{k}) &= \mathsf{A}(\boldsymbol{k}) - \frac{2 K^2 S^2}{h}
 \left(
\begin{array}{cc}
  \gamma^*_{1,\boldsymbol{k}} \gamma^{\phantom{*}}_{1,\boldsymbol{k}} & 0 \\
 0 &  \gamma^*_{2,\boldsymbol{k}} \gamma^{\phantom{*}}_{2,\boldsymbol{k}} \\
\end{array}
\right), \\
\mathsf{B}_{\text{eff}}(\boldsymbol{k}) &=  - \frac{ K (3 J \!+\! 2 K ) S^2}{h} 
\left(
\begin{array}{cc}
\gamma^{\phantom{*}}_{0,\boldsymbol{k}} \gamma^*_{1,\boldsymbol{k}}+ \gamma^*_{0,\boldsymbol{k}} \gamma^{\phantom{*}}_{2,\boldsymbol{k}}  & - 2 \gamma^*_{1,\boldsymbol{k}} \\
  - 2 \gamma^{\phantom{*}}_{2,\boldsymbol{k}} & 
\gamma^{\phantom{*}}_{0,\boldsymbol{k}} \gamma^*_{1,\boldsymbol{k}}+ \gamma^*_{0,\boldsymbol{k}} \gamma^{\phantom{*}}_{2,\boldsymbol{k}} \\
\end{array}
\right),
\end{align}
matrices. The canonical transformation generates an onsite correction and a second neighbor Dzyaloshinskii-Moriya term $\propto K^2S^2/h$ in the $\mathsf{A}_{\text{eff}}(\boldsymbol{k})$. The anomalous term $\mathsf{B}_{\text{eff}}(\boldsymbol{k})$ is $1/h$ and can be safely neglected in high fields, so the problem reduces to the diagonalization of the $\mathsf{A}_{\text{eff}}(\boldsymbol{k})$ $2\times 2$ matrix. The energies of the magnon excitations of the $4\times4$ and the effective $2\times2$ problem are identical including the $1/h$ corrections.

We note in passing that the matrix corresponding to the commutator of operators $[\mathcal{W}, \mathcal{H}]$ is 
\begin{equation}
    \mathsf{W}\cdot \eta \cdot  \mathsf{H} - \mathsf{H}\cdot \eta \cdot  \mathsf{W} \;.
\end{equation}

\subsection{Chern Number}

The $\mathsf{A}_{\text{eff}}(\boldsymbol{k})$  we need to diagonalize can be expressed as
 \begin{align} 
 \mathsf{A}_{\text{eff}} =& d_0(\boldsymbol{k}) \bm{1} + \frac{1}{2}\mathbf{d(\boldsymbol{k})}\cdot \bm{\sigma} \;,
 \end{align}
 where
 \begin{equation} 
 d_0(\boldsymbol{k}) =  h - (3 J +2 K) S -\frac{K^2 S^2}{h} \left( \gamma^*_{1,\boldsymbol{k}} \gamma^{\phantom{*}}_{1,\boldsymbol{k}} + \gamma^*_{2,\boldsymbol{k}} \gamma^{\phantom{*}}_{2,\boldsymbol{k}}\right), 
 \end{equation}
and
 \begin{equation} 
 \mathbf{d}(\boldsymbol{k}) = 
 \left(
\begin{array}{c}
 (3 J +2 K ) S (\gamma^*_{0,\boldsymbol{k}} + \gamma^{\phantom{*}}_{0,\boldsymbol{k}} ) \\
 i (3 J +2 K ) S (\gamma^*_{0,\boldsymbol{k}} - \gamma^{\phantom{*}}_{0,\boldsymbol{k}} ) \\
   -\frac{2 K^2 S^2}{h} \left( \gamma^*_{1,\boldsymbol{k}} \gamma^{\phantom{*}}_{1,\boldsymbol{k}} - \gamma^*_{2,\boldsymbol{k}} \gamma^{\phantom{*}}_{2,\boldsymbol{k}}\right) \\
\end{array}
\right) .
 \end{equation}
The $\bm{\sigma}=(\sigma^x,\sigma^y,\sigma^z)$ is a vector of the Pauli matrices, so
the $\mathbf{d}(\boldsymbol{k})$ acts as a fictitious magnetic field in the Brillouin zone. At each $\boldsymbol{k}$ we have the eigenvalues 
\begin{equation}
 \omega_{\pm}(\boldsymbol{k})= d_0(\boldsymbol{k}) \pm \frac{1}{2} d(\boldsymbol{k}) \;,
\end{equation} 
where $d(\boldsymbol{k})=|\mathbf{d} (\boldsymbol{k})|$. Each of these eigenvalues  forms a band in the Brillouin zone, with the spacing between the bands given by $d(\boldsymbol{k})$. The two bands can only touch when $|d(\boldsymbol{k})|=0$, which happens if $K=0$ or $3J+2K=0$.

The Berry curvature for the $2\times 2$ problem is given by
\begin{align}
F^{xy}_{\pm} (\boldsymbol{k}) 
&=  \pm \frac{i}{2} 
\frac{\mathbf{d}(\boldsymbol{k})}{d(\boldsymbol{k})^3} \!\cdot\! \left( \frac{\partial \mathbf{d}(\boldsymbol{k})}{\partial k_y} \!\times\! \frac{\partial \mathbf{d}(\boldsymbol{k})}{\partial k_x} \right) 
\label{eq:SFddd}\\
&=  \pm \frac{i}{2}
\mathbf{\hat{d}}(\boldsymbol{k}) \!\cdot\! \left( \frac{\partial \mathbf{\hat{d}}(\boldsymbol{k})}{\partial k_y} \!\times\! \frac{\partial \mathbf{\hat{d}}(\boldsymbol{k})}{\partial k_x} \right) \;, 
\label{eq:SFhdhdhd}
\end{align}
where $\mathbf{\hat{d}}(\boldsymbol{k})=\mathbf{d}(\boldsymbol{k})/d(\boldsymbol{k})$ is a unit vector. The Chern number of the band $\pm$ is then 
\begin{align}
C_{\pm} = \frac{1}{2\pi i} \int_{\text{BZ}} \! d{k_x} d{k_y} \; F^{xy}_{\pm} = \pm N_{s} \;,
\end{align}
where $N_s$ measures the number of skyrmions (topological defects) in the $\mathbf{d}$ field, as it follows from Eq.~(\ref{eq:SFhdhdhd}).

In the case of the Kitaev-Heisenberg model, introducing the notation
\begin{align}
u_c(\boldsymbol{k}) &= 
  \cos \boldsymbol{k}\!\cdot\!(\bm{\delta}_{z}\!-\!\bm{\delta}_{y}) 
+ \cos \boldsymbol{k}\!\cdot\!(\bm{\delta}_{x}\!-\!\bm{\delta}_{z}) 
+ \cos \boldsymbol{k}\!\cdot\!(\bm{\delta}_{y}\!-\!\bm{\delta}_{x}),
\nonumber\\
u_s(\boldsymbol{k}) &= \sin\boldsymbol{k}\!\cdot\!(\bm{\delta}_{z}\!-\!\bm{\delta}_{y})
 +\sin \boldsymbol{k}\!\cdot\!(\bm{\delta}_{z}\!-\!\bm{\delta}_{y})
 +\sin \boldsymbol{k}\!\cdot\!(\bm{\delta}_{z}\!-\!\bm{\delta}_{y}),
\end{align}
the expressions in the Berry curvature, Eq.~(\ref{eq:SFddd}), are 
\begin{align}
\mathbf{d}\!\cdot\!\left( \frac{\partial \mathbf{d}}{\partial k_y}\!\times\!\frac{\partial \mathbf{d}}{\partial k_x} \right) &=-  \frac{4(3 J \!+\! 2 K)^2 K^2 S^4}{27h} \left[(3 \!-\! u_c(\boldsymbol{k}))^2 \!-\! u_s^2(\boldsymbol{k}) \right] ,\label{eq:ddd}\\
d^2 &= \frac{4}{9} (3 J \!+\! 2 K)^2 S^2 (3+2 u_c(\boldsymbol{k}) )+\frac{16}{27} \frac{K^4}{h^2} u_s^2(\boldsymbol{k}) .
\end{align}
The triple product, Eq.~(\ref{eq:ddd}), is negative semidefinite in the whole Brillouin zone (it is 0 for $\boldsymbol{k} = 0$). The Chern number is therefore always finite, apart from the cases when $K=0$ (the  Heisenberg model), and when $3 J + 2 K = 0$. 
 In large fields, the $F^{xy}_{\pm} (\boldsymbol{k})$ is strongly peaked at the $K$ points, the corners of the hexagonal Brillouin zone.

\subsection{Thermal Hall conductivity}

In this section, we apply the expression for thermal conductivity 
\begin{align}
 \kappa^{xy} &=   \frac{1}{\beta} \sum_{n=\pm}\int_{\text{BZ}} d^2\boldsymbol{k} \; c_2(\rho_n)  \frac{F^{xy}_{n}(\boldsymbol{k})}{i} \;,
  \label{eq:kappa}
\end{align}
derived by Matsumoto et al. \cite{matsumoto2011rotational}, to the case of the Kitaev model. $\beta=1/ T$ is the inverse temperature and
\begin{align}
  \rho_n &= \frac{1}{e^{\omega_n\beta}-1}\;, \nonumber\\
  c_2(\rho) &= \int_{0}^{\rho}d t \; \ln^2(1+t^{-1})\;.
\end{align}
 Since the upper and lower bands have Berry curvatures with opposite signs, $ F_{+}^{xy}(\boldsymbol{k}) = -F_{-}^{xy}(\boldsymbol{k}) $, the expression for the thermal Hall effect simplifies to 
\begin{align}
 \kappa^{xy} &=  \int_{\text{BZ}} d^2\boldsymbol{k} \; \frac{c_2(\rho_{+})-c_2(\rho_{-})}{\beta}  
\frac{\mathbf{d}(\boldsymbol{k})}{d(\boldsymbol{k})^3} 
 \cdot \left( \frac{\partial \mathbf{d}(\boldsymbol{k})}{\partial k_y} \times \frac{\partial \mathbf{d}(\boldsymbol{k})}{\partial k_x} \right)  \;.
 \label{eq:kappaxydiff}
\end{align}
Since in large magnetic field the band dispersions and splittings are much smaller than the gap between the bands and the ground state, $\omega_{+}-\omega_{-}\ll d_0$, we expand Eq.~(\ref{eq:kappaxydiff}) in $d/d_{0}$: 
\begin{align}
 c_2(\rho_{+})-c_2(\rho_{-}) &= \int_{\rho_{-}}^{\rho_{+}}d t \; \ln^2(1+t^{-1}),
 \nonumber\\
 &\approx \left(\rho_{+}-\rho_{-}\right)  (d_0\beta)^2 \;,
\end{align}
where the difference of Bose occupation numbers is
\begin{align}
\rho_{+}-\rho_{-} &= 
-\frac{d \beta}{2\sinh^2(\frac{d_0\beta}{2})}+O\left(d^3\right)\;,
\end{align}
so that 
\begin{align}
\frac{1}{\beta} \left[c_2(\rho_{+})-c_2(\rho_{-})\right]  &\approx -\frac{ (d_0\beta)^2}{2\sinh^2(\frac{d_0\beta}{2})} d \;.
\end{align}
Eventually, we get the following simple expression for the thermal Hall conductivity:
\begin{equation}
  \kappa^{xy} = R(d_0\beta) \kappa^{xy}_\infty \;,
\end{equation}
where  
\begin{equation}
R(x) = \left(\frac{x}{2 \sinh \frac{x}{2}}\right)^2
\end{equation}
and
\begin{align}
\kappa^{xy}_\infty &=  \int_{\text{BZ}} d^2\boldsymbol{k}\;  
\frac{2}{d(\boldsymbol{k})^2}\mathbf{d}(\boldsymbol{k})\cdot \left( \frac{\partial \mathbf{d}(\boldsymbol{k})}{\partial k_x} \times \frac{\partial \mathbf{d}(\boldsymbol{k})}{\partial k_y} \right), \nonumber\\
& = 16 \pi \frac{S^2 K^2}{h}  \ln \frac{c K^2 S}{h |3 J + 2 K|} + \cdots 
\end{align}
in the leading order in $1/h$, where $c$ is a constant of order unity.
The temperature dependence stems purely from $R(d_0\beta)$. At low temperatures, the temperature dependence is thermally activated, while at high temperatures  $R\rightarrow 1$ and the conductivity saturates, with $\kappa^{xy}_\infty$ being the high temperature value.

In Fig.~$4$ in the main text, we show the low temperature thermal Hall conductivity as computed from Eq.~\ref{eq:kappa}.

\section{Non-linear spin-wave theory}
\newcommand{\mf}[1]{{#1}_{\textrm{eff}}}
In this section we outline the calculation of the dynamical structure
factor in non-linear spin-wave theory.  The starting point is the
Holstein-Primakoff expansion~\cite{holstein1940}, organized in powers
of $1/S$ (factoring out the overall $S^2$ scaling). Linear spin wave
theory appears when truncating to $O(1/S)$. To go to $O(1/S^2)$, one
must consider the effects of magnon-magnon interactions, including
three- and four-body terms in the Holstein-Primakoff bosons (magnons).

The dynamical structure factor at $O(1/S^2)$ requires the computation
of the magnon Green's function as well as several higher order
dynamical correlation functions.  It is useful to consider three
distinct pieces: the transverse-transverse part which involves only
the magnon Green's function, $\mat{\mathsf{G}}(\vec{k},\omega)$, the
transverse-longitudinal parts which involve three-magnon correlation
functions and the longitudinal-longitudinal parts which involve
four-magnon correlation functions~\cite{mourigal2013dynamical}.  While
the transverse-transverse part has $O(1/S)$ contributions, the other
two parts appear first at $O(1/S^2)$. We note that the Green's
function also appears in the transverse-longitudinal part of the
dynamical structure factor, while the longitudinal-longitudinal part
involves only the free magnon Green's function at leading
order~\cite{mourigal2013dynamical}.

Typically, the transverse-longitudinal and longitudinal-longitudinal
parts are small relative to the leading transverse-transverse
contributions.  The central ingredient is then (retarded) magnon
Green's function~\cite{blaizot1986quantum}
\begin{equation}
  \mat{\mathsf{G}}(\vec{k},\omega) = \left[(\omega+i0^+)\mat{\eta} - \mat{\mathsf{M}}(\vec{k}) - \mat{\Sigma}_{\mat{\mathsf{M}}}(\vec{k},\omega)\right]^{-1},
\end{equation}
where $\mat{\mathsf{M}}(\vec{k})$ is the linear magnon dispersion
matrix (see Eq.~($2$) of the main text) and $\mat{\eta} =
{\textrm{diag}}(+\mat{1},-\mat{1})$ is due to the bosonic Bogobiulov
transformation~\cite{blaizot1986quantum}. The self-energy,
$\mat{\Sigma}_{\mat{\mathsf{M}}}(\vec{k},\omega)$, appears due to the
magnon-magnon interactions and can be evaluated perturbatively in
powers of $1/S$, starting from the solution of the linear spin-wave
problem encoded in $\mat{\mathsf{M}}(\vec{k})$. This Green's function
and the self-energy are both matrices with sublattice indices and have
both normal and anomalous contributions~\cite{blaizot1986quantum,chubukov1994large}.

We identify two distinct types of contributions to the self-energy:
static (frequency independent) and dynamic (frequency dependent).  The
static contributions arise from Hartree-Fock-like diagrams involving
the four-magnon interactions as well as (in principle) tadpole-like
diagrams arising from the three-magnon interaction. The dynamic
contributions arise purely from the three-magnon interactions.  In
addition to renormalizing the one-magnon spectrum they are also
responsible for magnon
decay~\cite{chubukov1994large,chernyshev2009triangular}, possibly endowing the
one-magnon states with finite lifetimes.

The Holstein-Primakoff expansion is formally controlled in $1/S$, and
is thus a systematic approximation scheme when $S \gg
1$. Alternatively, it can be viewed as an expansion in the magnon
density $\rho \equiv \braket{a^{\dagger} a^{}}/(2S)$, and is
controlled in the limit, $\rho \ll 1$. For arbitrary $S$ this limit
can be reached systematically through the application of a large
magnetic field. For small $S$ or for small fields however its validity
is more limited. Carried to order $O(1/S^2)$ two key issues are
apparent: (a) it is confined by the classical phase boundaries and (b)
the two-magnon spectrum does not reflect the renormalization of the
one-magnon spectrum due to interactions. While one could alleviate some
of these issues by proceeding to higher order in $1/S$, the technical
complexity of such calculations is prohibitive both computationally and
analytically.

Instead, we adopt a self-consistent approach, allowing the static part
the Green's function to renormalize the linear spin-wave dispersion.
Specifically, we introduce a renormalization
\begin{equation}
    \label{eq:green}
  \mat{\mathsf{G}}(\vec{k},\omega) = \left[(\omega+i0^+)\mat{\eta} - \mf{\mat{\mathsf{M}}}(\vec{k}) - (\mat{\Sigma}_{\mf{\mat{\mathsf{M}}}}(\vec{k},\omega)-\delta \mat{\mathsf{M}}(\vec{k}))\right]^{-1},
\end{equation}
where $\mf{\mat{\mathsf{M}}}(\vec{k}) \equiv \mat{\mathsf{M}}(\vec{k})
+\delta \mat{\mathsf{M}}(\vec{k})$ and we take (formally) $\delta
\mat{\mathsf{M}}(\vec{k}) \sim O(1/S^2)$. Note that the self-energy is
evaluated using the energies and eigenvectors associated with
renormalized free problem, $\mf{\mat{\mathsf{M}}}(\vec{k})$, not the
original $\mat{\mathsf{M}}(\vec{k})$.  The renormalization, $\delta
\mat{\mathsf{M}}(\vec{k})$, is then chosen to cancel the static,
Hartree-Fock-like contributions to the self-energy. This procedure
does not not strictly include only $O(1/S^2)$ contributions and is thus in
some sense uncontrolled. However, using such an approach we can
account for some of the change in the one-magnon energies due to
interactions in the free problem, as well as access regions of the
phase diagram outside the usual classical phase boundaries, without
having to go to higher order in $1/S$.

Applied to the problem at hand, for the bulk case we assume the system is in
the fully field polarized state with the magnetic moments aligned with the
applied $[111]$ field. Due to the anisotropic Kitaev exchange, even in
this colinear state there are both three- and four-magnon
interactions. The three-magnon interactions generically induce
spontaneous decay of the one-magnon excitations when they overlap with
the two-magnon continuum~\cite{chernyshev2013colloquium}.

Our implementation considers a finite system of size $N = 2L^2$. We
first solve for $\delta \mat{\mathsf{M}}(\vec{k})$ for each
wave-vector through self-consistent iteration (terminating when the
maximum change in the correction is $\lesssim 10^{-8}$).  In the
classically allowed regions we can initialize the iteration trivially
with $\delta \mat{\mathsf{M}}(\vec{k}) = 0$. However, to access the
critical field $h_c=2$ for $K=+1$, we begin with small chemical
potential $\delta \mat{\mathsf{M}}(\vec{k}) = \mu \mat{1}$ where $\mu
\sim 0.1$ to avoid the classical instability (the final result is
independent of the choice of $\mu$).  Once
$\mf{\mat{\mathsf{M}}}(\vec{k})$ is determined for each wave-vector,
we then compute $\mat{\Sigma}_{\mf{\mat{\mathsf{M}}}}(\vec{k},\omega)$
on a fine grid in frequency, including a small width $0^+ \rightarrow
10^{-3}$ to resolve any singularities. Performing the inversion in
Eq.~(\ref{eq:green}) numerically (including again a small width), we
then obtain $\mat{\mathsf{G}}( \vec{k},\omega)$ which determines the
dominant transverse-transverse part of the dynamical structure
factor. The remaining free Green's functions and sums involved
in the remaining parts~\cite{mourigal2013dynamical} are evaluated
similarly. The results in the main text show the full dynamical
structure factor $S(\vec{k},\omega)$ (as given in Eq.~($3$)
of the main text) including the transverse-transverse,
transverse-longitudinal and longitudinal-longitudinal contributions.

For the case with open boundaries we consider systems of $N = 2LW$
sites where $W=5$, following the same strategy to evaluate the
dynamical structure factor as in the bulk case . The main modification
necessary arises at the classical level from the presence of open
boundaries. Since the spins at the boundaries have fewer neighbors
than those in the bulk, the classical ground state is no longer
uniform, with the moment direction deviating from $[111]$ as one
approaches the edges. Due to the lower symmetry, the static tadpole
diagrams are non-zero and are included in the self-consistent
iteration described above. The presence of these diagrams implies that
the one-magnon expectation values do not vanish and thus there is
finite $O(1/S^2)$ correction to the canting of the moments away from
$[111]$. In addition, due to the imposition of open boundaries, the
classical critical field is also lowered, with $h_c < 2$ for
$K=+1$. As for the bulk case, the results in the main text show the
layer-resolved dynamical structure factor, including the
transverse-transverse, transverse-longitudinal and longitudinal
contributions.  To compare directly with the DMRG results, the
definition of the layer-dependent structure factor only includes 
$S^{xx}(\vec{k},\omega)$, as described in Fig.~$3$ of the main text.

\section{Bulk Dynamical Structure Factors}

\begin{figure}[t!]
\subfloat{%
  \includegraphics[width=\linewidth]{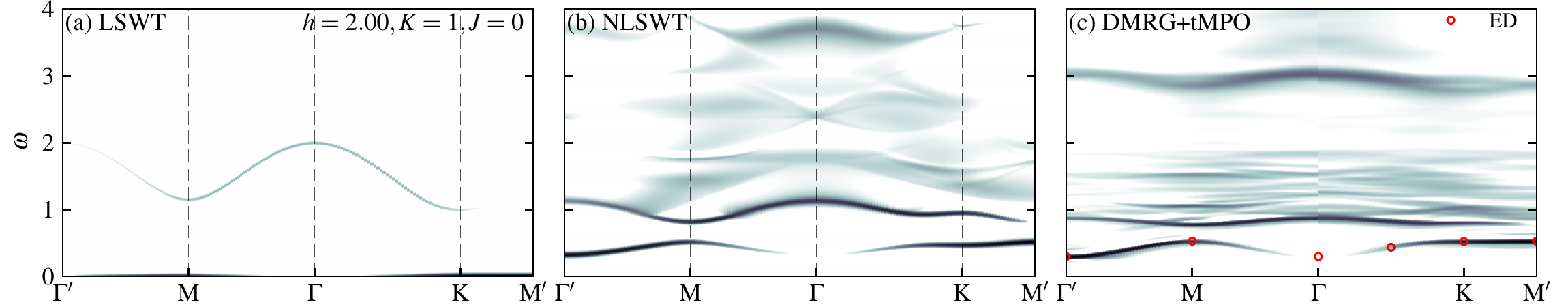}%
}

\subfloat{%
  \includegraphics[width=\linewidth]{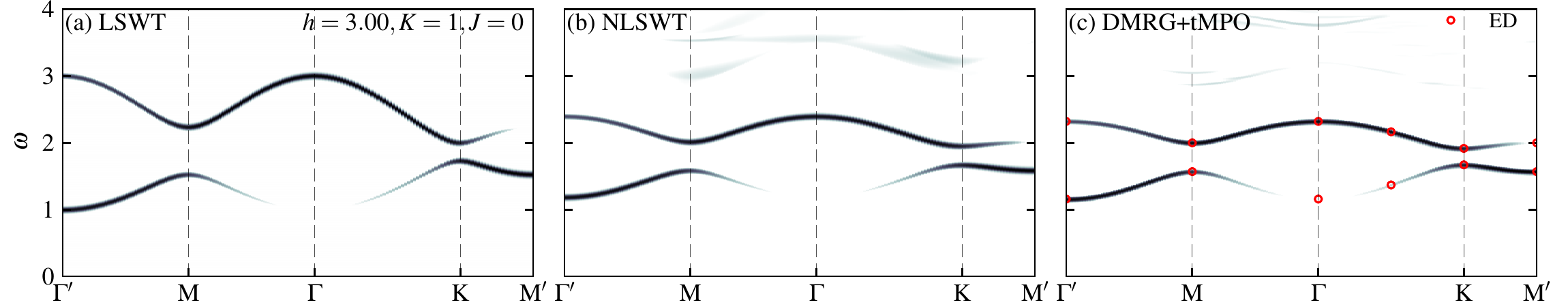}%
}

\subfloat{%
  \includegraphics[width=\linewidth]{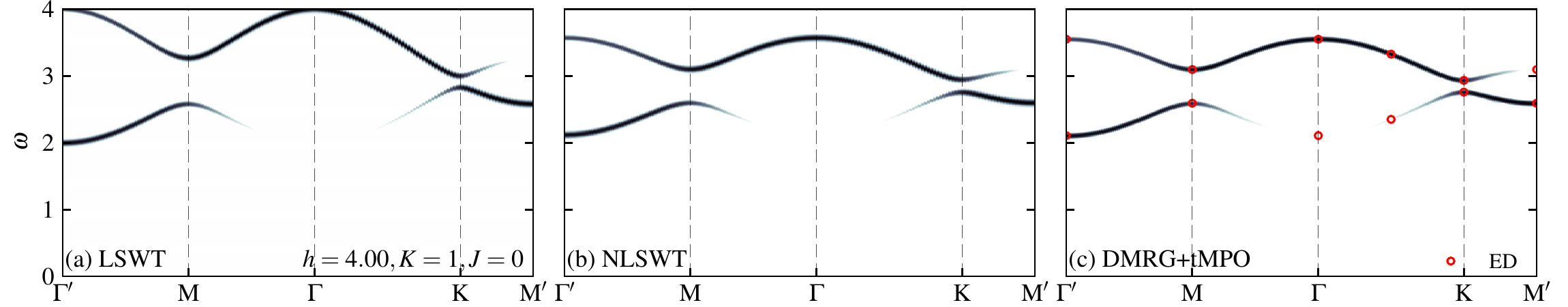}%
}

\caption{Dynamical structure factor at the antiferromagnetic Kitaev point $\vartheta=\pi/2$ at fields $h=2, 3, 4$ from top to bottom. The $h=3$ figure appears also in the main text. The intensity scale is logarithmic from $5\times 10^-3$ to $1$.  At each field, results are shown for linear spin wave theory (left), nonlinear spin wave theory (middle) and time dependent DMRG (right). The red points in the right-hand plots are exact diagonalization results for the symmetric $24$ site cluster.}
\label{fig:AFMKit}
\end{figure}

Fig.~\ref{fig:AFMKit} shows the dynamical structure factors for $h=2,3,4$ at the antiferromagnetic Kitaev point using linear spin wave theory, interacting spin wave theory to $1/S^2$ and time dependent DMRG along high symmetry lines. The results for $h=3$ are shown in the main text. Linear spin wave theory disagrees significantly at all these fields - the bandwidth is overestimated - while interacting spin wave theory to $1/S^2$ agrees very well at $h=3$ and $h=4$ with the DMRG. At $h=2$ the lowest magnon band is a flat band at zero energy (the calculation here is actually for $h=2+\epsilon$) while the fully polarized state is stable in the interacting spin wave calculation. However, at $h=2$ the $1/S^2$ and DMRG calculations are mutually inconsistent in both the single magnons and the higher energy continuum scattering. Exact diagonalization results for the $24$ site hexagonal cluster in the single magnon sector agree very well with the DMRG. 

A similarly organized set of figures are shown for the ferromagnetic Kitaev point at $h=1,2$. While linear spin wave theory has the symmetry explained in Section~\ref{sec:map}, matching the spectra at the ferromagnetic and antiferromagnetic Kitaev points, this mapping breaks down in the presence of interactions and, indeed, the $\vartheta=\pi/2$ and $3\pi/2$ results are dramatically different. In particular, a multiparticle continuum visibly overlaps the upper single magnon bands in the DMRG causing considerable broadening. This is partially captured within interacting spin wave theory. 

We draw the attention of the reader to the high intensity broad and nearly flat intensity visible at $h=2$ in the DMRG around $\omega=3$ (panel~\ref{fig:AFMKit}(c) in the top row). This feature of the multimagnon intensity persists to higher fields. The precise nature of this object is a question that we leave for future work.

\begin{figure}[t!]
\subfloat{%
  \includegraphics[width=\linewidth]{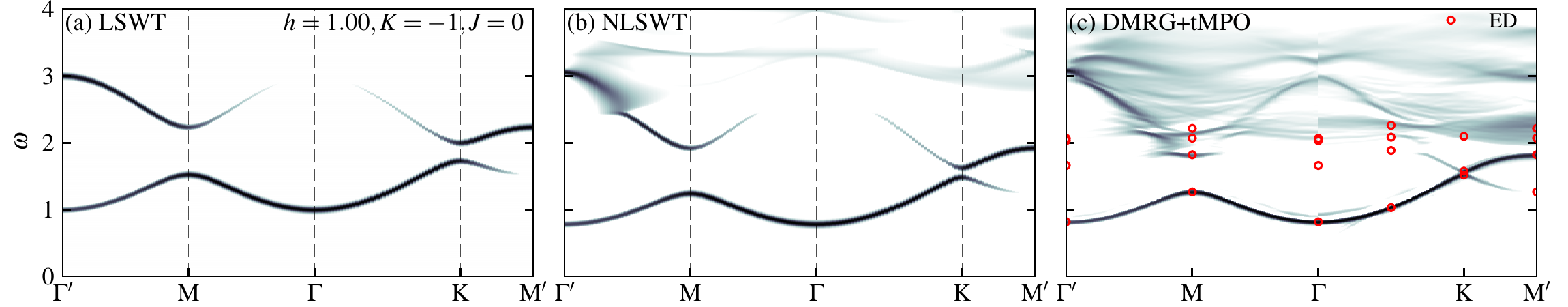}%
}

\subfloat{%
  \includegraphics[width=\linewidth]{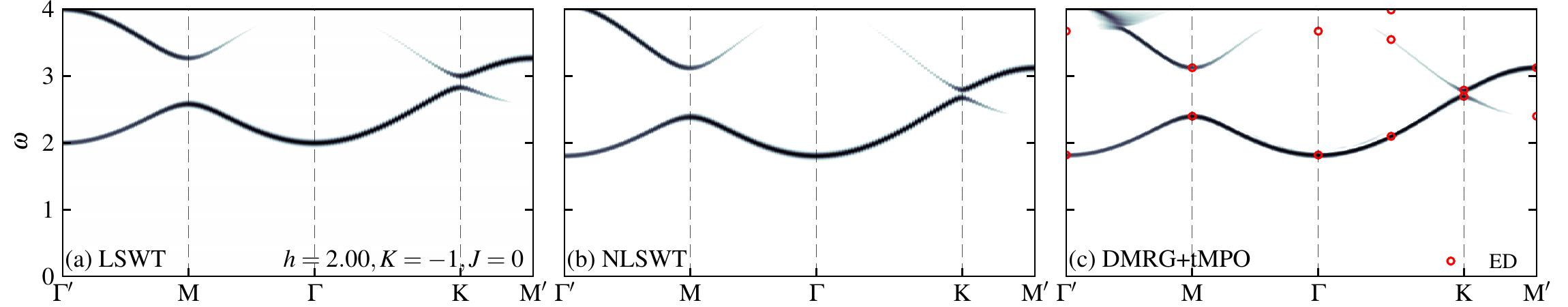}%
}

\caption{Dynamical structure factor at the ferromagnetic Kitaev point $\vartheta=3\pi/2$ at fields $h=1, 2$ from top to bottom. The intensity scale is logarithmic from $5\times 10^-3$ to $1$.  At each field, results are shown for linear spin wave theory (left), nonlinear spin wave theory (middle) and time dependent DMRG (right).}
\label{fig:FMKit}
\end{figure}

\section{Dynamical Correlations in the Slab Geometry}

Here we report further slab geometry results for the dynamical correlations as computed using linear spin wave theory, interacting spin wave theory and time dependent DMRG. The slab is described in the main text and illustrated at the bottom of Fig.~$3$ there. 

\begin{figure}[t!]
\subfloat{%
  \includegraphics[width=0.7\linewidth]{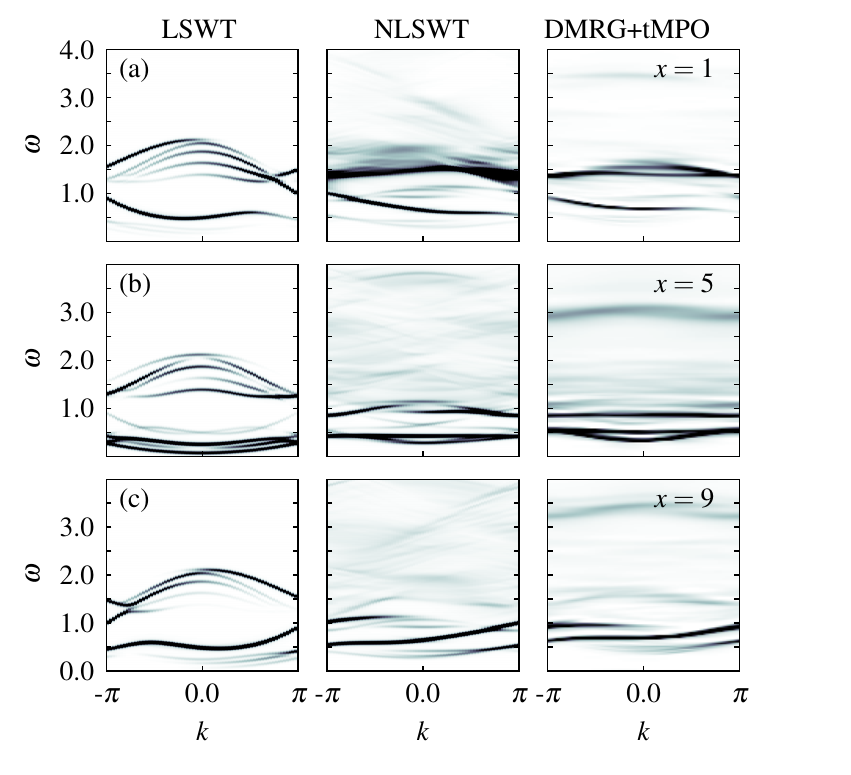}%
}

\subfloat{%
  \includegraphics[width=0.7\linewidth]{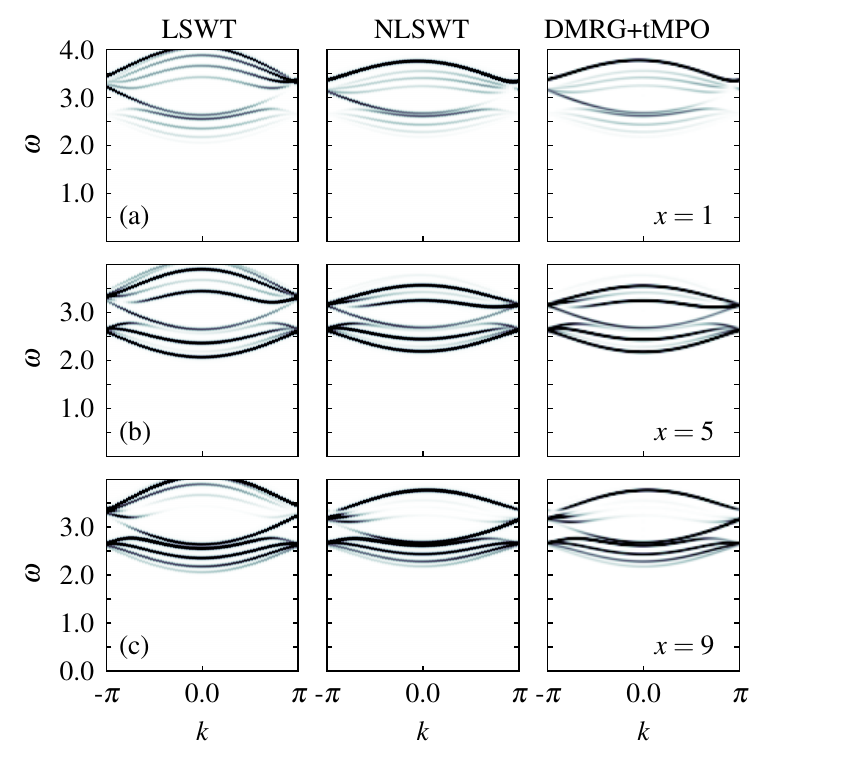}%
}

\caption{Dynamical correlations on a slab geometry at the antiferromagnetic Kitaev point $\vartheta=\pi/2$ at fields $h=2, 4$ from top to bottom. At each field, results are shown for linear spin wave theory (left), nonlinear spin wave theory (middle) and time dependent DMRG (right).}
\label{fig:AFMSlab}
\end{figure}

Fig.~\ref{fig:AFMSlab} shows results at the antiferromagnetic Kitaev point at $h=2$ (top) and $h=4$ (bottom). The case of $h=3$ is given in the main text. At $h=4$ the finite size progenitor of the chiral mode on the semi-infinite slab is clearly visible in panel (a)(left) running between the bulk bands with highest intensity to the left of that panel. The principal result of including interactions is a narrowing of the bandwidth: the same mode being visible in (b) and (c). The middle of the slab has visible but lower intensity between those bands that persist into the bulk geometry. The lower set of panels again show the chiral mode - this time on the right of each panel. At $h=2$, interactions play a much larger role because the lower threshold to two magnon states begins within the upper single magnon bands. The result is that there is considerable broadening of the upper single magnon modes. Nevertheless, the edge state appears to survive the presence of interactions.

\begin{figure}[t!]
\subfloat{%
  \includegraphics[width=0.7\linewidth]{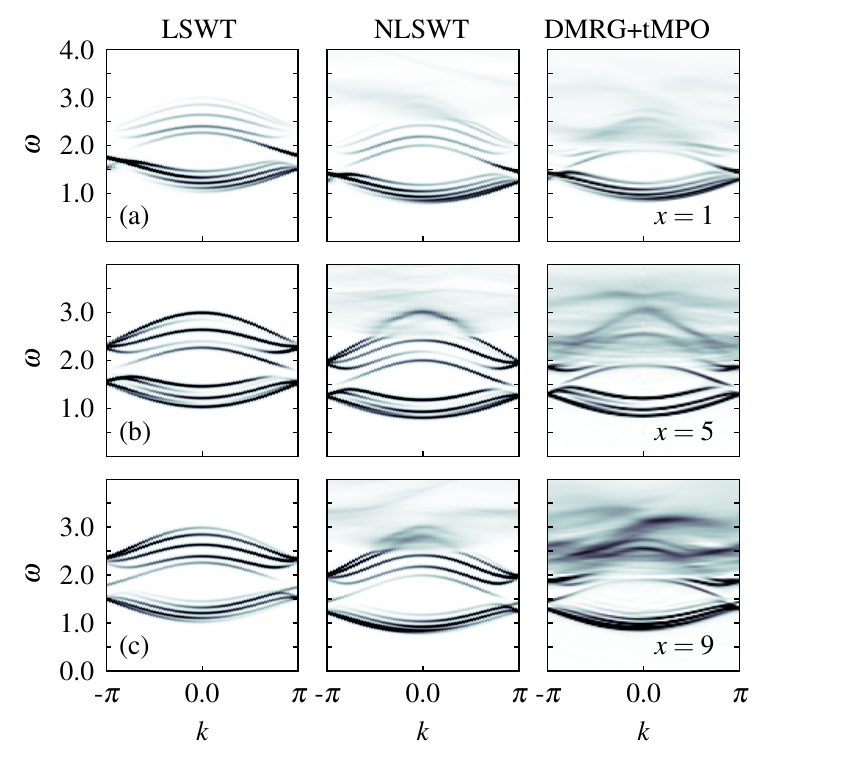}%
}

\subfloat{%
  \includegraphics[width=0.7\linewidth]{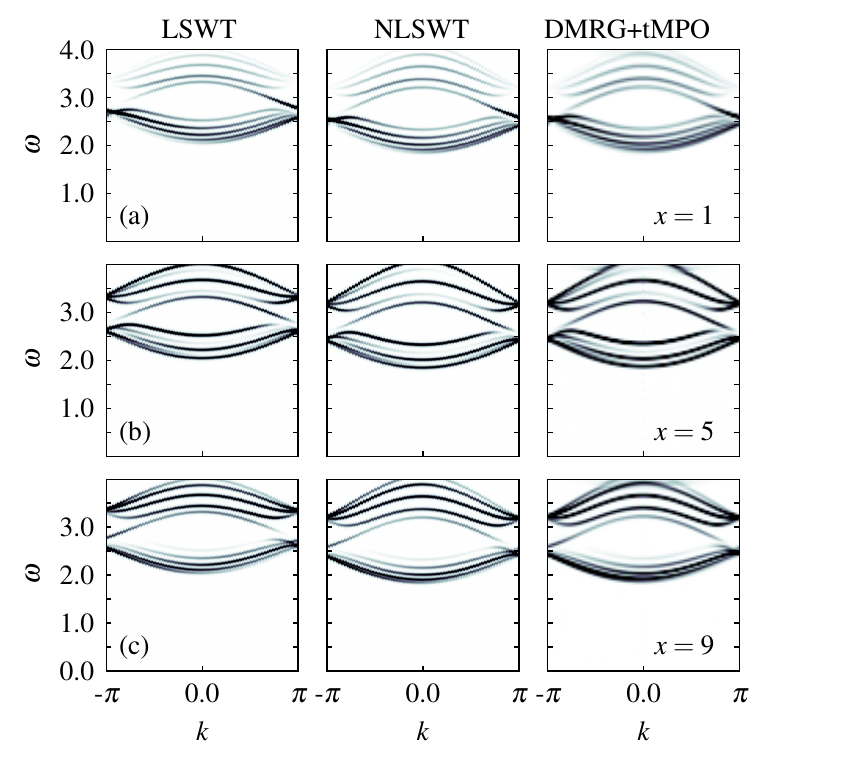}%
}

\caption{Dynamical correlations on a slab geometry at the ferromagnetic Kitaev point $\vartheta=\pi/2$ at fields $h=2, 4$ from top to bottom. At each field, results are shown for linear spin wave theory (left), nonlinear spin wave theory (middle) and time dependent DMRG (right).}
\label{fig:FMSlab}
\end{figure}

The symmetry between noninteracting magnon spectra in the bulk under $\vartheta\rightarrow\vartheta+\pi$ ceases to hold on the slab geometry because the presence of an edge causes the ground states to be affected by the change in the coupling. We find that the chiral edge mode on the open geometry is not as clearly visible in the ferromagnetic Kitaev model as it is in the antiferromagnetic case. In Fig.~\ref{fig:FMSlab} the chiral mode appears as inter-bulk band intensity in panel (a) with the same sign of the velocity as in (a) of Fig.~\ref{fig:AFMSlab}. At $h=2$, interactions bring about a fairly mild renormalization of the bands while, at $h=1$, the multimagnon continua are clearly visible. The upper bulk single magnon modes are destroyed through coupling to these additional states while the whole block of single magnon states is pushed to lower energies. Despite this dramatic effect of interactions, once again, the edge mode appears to survive.

\end{document}